# Can x2vec Save Lives?

*Integrating Graph and Language Embeddings for Automatic Mental Health Classification*

Alexander Ruch[1,2]

[1] Cornell University, `amr442@cornell.edu`
[2] Graphika Inc., `alex.ruch@graphika.com`

**Abstract**

Graph and language embedding models are becoming commonplace in large scale analyses given their ability to represent complex sparse data densely in low-dimensional space. Integrating these models' complementary relational and communicative data may be especially helpful if predicting rare events or classifying members of hidden populations – tasks requiring huge and sparse datasets for generalizable analyses. For example, due to social stigma and comorbidities, mental health support groups often form in amorphous online groups. Predicting suicidality among individuals in these settings using standard network analyses is prohibitive due to resource limits (e.g., memory), and adding auxiliary data like text to such models exacerbates complexity- and sparsity-related issues. Here, I show how merging graph and language embedding models (`metapath2vec` and `doc2vec`) avoids these limits and extracts unsupervised clustering data without domain expertise or feature engineering. Graph and language distances to a suicide support group have little correlation ($\rho < 0.23$), implying the two models are not embedding redundant information. When used separately to predict suicidality among individuals, graph and language data generate relatively accurate results (69% and 76%, respectively); however, when integrated, both data produce highly accurate predictions (90%, with 10% false-positives and 12% false-negatives). Visualizing graph embeddings annotated with predictions of potentially suicidal individuals shows the integrated model could classify such individuals even if they are positioned far from the support group. These results extend research on the importance of simultaneously analyzing behavior and language in massive networks and efforts to integrate embedding models for different kinds of data when predicting and classifying, particularly when they involve rare events.

## 1   Introduction

Complex social dynamics like behavior and communication patterns are often sparse in the sense that such data for any given individual, group, or entity have few observations of interest and many null or irrelevant observations. For example, in a network of 1,000,000 individuals, someone who has interacted with just 1% of the network has large number of ties to others (10,000), but their vector of actual and potential ties to others is nearly empty. The same outcome develops from a text that includes 100 unique words from a vocabulary of 10,000 words. This sparsity compounds the already complex practices of network analysis and natural language processing. Embedding models, however, are useful solutions to this problem, as they effectively (in terms of time and memory) learn dense low-dimensional representations of complex sparse data. In other words, embedding models can represent the aforementioned individual's network connectivity in a 128-dimension vector with non-null elements instead of a mostly empty 1,000,000-dimension vector.

Embedding models were first developed in natural language processing to represent similarities between words given the similar linguistic contexts in which they co-appear (e.g., `word2vec` in Mikolov et al. 2013). These models were then extended to embed words using their subword information (positioning "write" close to "writes" and "wrote") and extended to embed sentences and documents based on the words and topics in them (Bojanowski et al. 2017; Le and Mikolov 2014). Models to embed networks were based on these techniques, treating the nodes in a graph like words in a document. Nodes are sampled from the graph using random walks (Perozzi, Al-Rfou, and Skiena 2014), and the random walk algorithm can be adjusted to account for homophily and structural equivalence (Grover and Leskovec 2016) and for heterogeneous/multi-modal graph structure (Dong, Chawla, and Swami 2017).

Given their facility for analyzing complex network and language problems, integrating these two complementary embeddings may overcome the limitations of any single embedding and extend computational social science by making possible simultaneous analysis of multiple types of action. For example, integrated models may be especially helpful if one's research involves predicting rare events or classifying members of hidden populations, tasks requiring huge and sparse datasets for generalizable models. Using standard network analyses in these cases can be prohibitive due to resource limits (e.g., memory), and adding auxiliary data like text to complement these models further exacerbates their complexity and sparsity-related issues, as one sparse matrix combined with another often produces an even larger sparse matrix. For example, suicidality is stigmatized worldwide and frequently develops along with a wide range of other physical and mental health comorbidities (Corrigan 2004; Corrigan and Watson 2002). Individuals struggling with suicidality often seek sporadic support in amorphous online groups, making it difficult to classify members of these groups and predict those who may become suicidal. Producing representative models of these outcomes requires lots of sparse data, in other words, as most individuals never post in a support group and those who do may post in a large number of other online groups before they finally post in a support group.

In this paper, I show how integrating graph and language embedding models (`metapath2vec` and `doc2vec`) avoids the aforementioned limits while extracting unsupervised clustering data on networks and language without requiring domain expertise. I then measure how correlated these embeddings' distances are with each other to evaluate the extent to which they are extracting redundant similarity data. Next, I compare graph and language embeddings' separated and integrated abilities to generate accurate predictions of rare events: posting submissions to Reddit's *r/SuicideWatch* subreddit for suicide support in a heterogeneous network of 35.5 million nodes and 190 million edges. Finally, I quantitatively and qualitatively evaluate the integrated model's ability to avoid false-positives and false-negatives in this context. The results of this study extend research on the importance of simultaneously analyzing behavior and language in massive networks and efforts to integrate different kinds of embedding models for prediction and classification tasks, especially when they involve rare events.

All data used in this study are publicly available from Reddit. The code used to sample the network and perform graph and language embeddings, natural language processing, and machine learning predictions are available from the author upon request. Care is taken to avoid presenting personally identifiable information in this paper, and all examples of submissions are paraphrased. This work was approved by the appropriate Institutional Review Board (IRB) of the author. Please note that *this work does not make any diagnostic claims related to mental illness or suicide*.

## 2 Related Work

### 2.1 Network Analysis

Similarity between nodes in a network can be measured in various ways; however, many methods focus on extracting nodes' structural equivalence: the extent to which two or more nodes share the same set of edges to other nodes in the network (Borgatti and Grosser 2015). These nodes hold the same position in a network and are equivalent to each other even if they share no edge(s) between themselves. For example, if a company had two managers who worked with the same employees but who did not work with each other, these managers would have structurally equivalent positions in the company's network. Moreover, since these managers are connected to all of the same people, they can be said to be redundant in the network and may replace one another, as neither works or interacts with anyone else in the network who the other does not also work or interact. In many cases, however, we do not wish to know if two nodes share exactly the same set of neighbors.

Methods of measuring nodes' similarities can either derive continuous distances between nodes or identify classes of nodes that are more or less similar (e.g., for clustering or community detection). For example, continuous measures include the following methods. Euclidean distance equals the number of network neighbors that two nodes do not share and is normalized by dividing by the total number of nodes in the network. The Pearson correlation coefficient can also be used to get a normalized measure of shared network neighbors. The Jaccard coefficient equals the intersection of network neighbors shared divided by the union of network neighbors shared. Cosine similarity takes into consideration nodes' degree by taking the number of network neighbors nodes share and dividing it by the geometric mean of those nodes' degrees. Shi and Macy (2016) find that each of these methods are biased when measuring similarity among nodes in large networks, as the nodes in these networks can have highly skewed degree distributions which can overemphasize the effect of nodes' out-degrees on their similarity in many of these methods. They recommend using the standardized co-incident ratio in such cases, as its measure of similarity – based on the number of observed common neighbors divided by the expected number of common neighbors in a random network – is not biased in large networks with millions of nodes.

By contrast, similarity measures that identify discrete groups of similar nodes classify sets of nodes that share many connections within their respective groups but have little or no connection to other groups. Such groups are often called cohesive subgroups (Frank 1995, 1996). For example, using hierarchical agglomerative clustering on a network's similarity matrix (generated from any of the metrics in the previous paragraph) produces a tree-like structure of groups of increasingly similar nodes from pairs of nodes to small groups of nodes to larger groups of nodes all the way to where the whole network is clustered into one group. One can use a dendrogram to visualize this tree and select a cutoff threshold for similarity to generate classes of nodes that formed within the threshold. Modularity is a similar algorithm that maximizes the proportion of edges lying between nodes in a group minus the expected proportion of edges existing between nodes in a group at random. Unfortunately, this method can suffer from a "resolution limit" in which relatively small classes of similar nodes are overlooked by the algorithm (Fortunato and Barthé 2007).

Hierarchical stochastic block models (HSBMs) attempt to bridge the gap between continuous and discrete measures of similarity. In HSBMs, levels of increasingly similar groups (or "blocks") of nodes are produced using algorithms that iteratively partition nodes into groups in which member nodes share many connections to each other while sharing no or few connections to nodes in other groups. These models do not suffer from the "resolution limit" that hampers modularity measures (Peixoto 2014). Additionally, these models can adjust nodes' similarities based on their degree and can generate overlapping (i.e., continuous) groups as well. Lastly, HSBMs can be used to analyze core-periphery structures in networks and to predict whether edges exist between nodes. Despite these advances, HSBMs are computationally expensive and may be unsuccessful in large networks, as their algorithms require keeping a large amount of data in memory (Fortunato and Hric 2016).

Given the complexity of accurately measuring network similarity at scale, attempts to add auxiliary language data to network analyses have been limited – yet, such data is recognized as important to accurately capturing complex social interactions in ways that extend beyond what either behavior or communication data captures on its own (Benamara, Inkpen, and Taboada 2018). Many studies that have integrated these two forms of data have used qualitative discourse analysis of text, which – while powerful at capturing nuanced meanings in interactions – are limited to small scale studies and require expert working knowledge of the given context (Moser et al. 2013). On the other hand, many studies that have used quantitative text analysis approaches, like natural language processing, in addition to network analysis have often used few measures of network structure and similarity (Benamara et al. 2018; Gonzalez-Bailon et al. 2010). Bail's work (2016) contrasts with these trends by directly integrating language similarity data with network structure measures (e.g., closeness and betweenness centrality) in his study of autism spectrum disorder advocacy organizations and their communication practices. Though his method enabled the analysis of cultural bridges through a combined behavior and communication framework, his approach to natural language processing used a bag-of-words tokenization that removed much of the language's context that comes from how words are used in a particular order (e.g., "I know, vaccines don't cause autism" vs "I don't know, vaccines cause autism"). Leaving such properties of language out of our analyses may cause us to miss important dynamics in modeling social interaction and similarity.

## 2.2 Embedding Models

### 2.2.1 Graph Embedding

The similarity measures mentioned above run on networks' adjacency matrices. If a network has 1,000,000 nodes, the size of this matrix will be $1,000,000_2$ (i.e., it will contain one trillion elements). If the network is sparse, the elements in this adjacency matrix will be mostly empty, contributing to substantial computational overhead that generates little information. Embedding models avoid these limitations by learning dense low-dimensional representations of data. In other words, one can use embedding models to learn similarities between nodes in this network of 1,000,000 nodes without analyzing its adjacency matrix and produce a matrix of 1,000,000 rows (one for each node) and $D$ columns, where $D$ represents the depth of similarity dimensions representing the network (typically 64, 128, 256). The positional values of these $D$ elements will be rarely be empty, which addresses our sparsity concern, and the fact that $D$ is miniscule in size compared to the number of nodes in the network allows us to avoid memory limits. This embedding matrix can then be used for other downstream similarity tasks like cluster analysis.

Graph embeddings can be constructed through three methods, each of which has its own strengths and limits (Cai, Zheng, and Chen-Chuan Chang 2017; Goyal and Ferrara 2017). First, factorization of graph adjacency matrices (Ahmed et al. 2013) can extract similarities between nodes and their neighbors (first order proximity) and similarities between nodes and their neighbors' neighbors (second order proximity), as is done in `LINE` (Tang et al. 2015). This approach is limited, however, as it must run on networks' adjacency matrices, which hampers its ability to run on networks that are too large to load into memory. `HOPE` (Ou et al. 2016) attempts to overcome this limit by using Singular Value Decomposition to create a low-dimension representation of the adjacency matrix and then embeds first and second order proximities in the similarity matrix.

While `HOPE` does reduce memory load and embedding's computational complexity, factorization-based embedding methods still require all nodes in the network to be loaded into memory at once. Many deep learning embedding methods have the same requirement. For example, `SDNE` (Wang, Cui, and Zhu 2013) passes nodes and their neighbors through a series of autoencoders to learn the graph's structure. This method is both computationally efficient and capable of learning non-linear features of structural similarity without supervision; however, because the whole network must be loaded in memory at once, it cannot be used in cases where the target network cannot fit in memory or where it is not possible to collect the whole network (e.g., when nodes and edges may be missing or cannot be extracted due to sampling limitations or restrictions).

Using random walk sampling methods in deep learning embedding models avoids the space limit issue of models that require loading the whole network into memory. Many of these models learn nodes' primary and secondary similarities by iterating through sequences of nodes extracted from random samples and using a Skip-Gram-based model to generate embeddings that maximize the probability of observing the neighbors of nodes – up to a specific window length – conditional on nodes' present embeddings. For example, if a walk from a sample included the nodes *user$_1$*, *user$_2$*, *user$_3$*, *user$_4$*, *user$_5$*, and so on, and if nodes are embedded over a window of 2, then the Skip-Gram model would increase *user$_1$*'s similarity to *user$_2$* and *user$_3$*, then increase *user$_2$*'s similarity to *user$_1$*, *user$_3$*, and *user$_4$*, then increase *user$_3$*'s similarity to *user$_1$*, *user$_2$*, *user$_4$*, and *user$_5$* – and so on. To quicken training, the Skip-Gram model is often run jointly with a negative sampling process that adds distance between the focal node and a set of non-neighbor nodes. The details of this algorithm and how it is applied in the present paper are outlined in **Section 3.2**.

`DeepWalk` (Perozzi et al. 2014) was the first graph embedding model to use the Skip-Gram model. It used strictly random walks to generate samples of nodes to train the embedding model. Grover and Leskovec (2016) also use Skip-Gram to embed nodes, but they use two parameters to bias the random walk's probability of returning to the node last sampled and to bias the walk's probability of sampling nodes close to the node last sampled versus that are nodes far from it. Together with the Skip-Gram model, this biased random walk sampling process helps embed nodes that can have more or less homophily and structural equivalence. The `metapath2vec` model (Dong et al. 2017) extends this approach for embedding heterogeneous graphs comprised of different types of nodes (e.g., universities, researchers, papers, and conferences). To do this, one defines a metapath-biased random walk scheme: a set of node types to sample in a specific order (e.g., {researcher, paper, conference, paper, researcher} embeds the similarities of researchers based on the conferences in which they both present papers). A variant, `metapath2vec++`, modifies the negative sampling procedure as well to generate a set of embeddings that are specific to each type of node encountered (e.g., researchers are embedded near researchers and far from papers, which are embedded closely).

### 2.2.2 Language Embedding

The Skip-Gram model and negative sampling process were originally developed to embed words using `word2vec` (Mikolov et al. 2013). In the context of language, this embedding model learns the maximize the probability of words that co-occur. For example, if you read the sentence "I drink ____ for breakfast," words like coffee, tea, milk, juice, and water probably come to mind based on how you have likely heard or read this same sentence with each of these words in the middle many times before. In other words, each of these breakfast beverages shares a similar language context. One limitation of this model, however, is that it does not embed information about word order. For example, it cannot estimate the meaning and similarity of "Buffalo wings" to other words by simply combining the embeddings of "buffalo" and "wings."[1] The `word2vec` sampling algorithm can be updated, however, to identify and embed such phrases, however (Mikolov et al. 2013). This approach has since been modified to embed subword information (Bojanowski et al. 2017) as well as complete sentences and documents (Le and Mikolov 2014). The latter method of embedding documents with `doc2vec` (Le and Mikolov 2014) is particularly useful, as it is able to learn to represent semantic as well as syntactic information about language. The details of the `doc2vec` algorithm and how its two variations are applied in the present paper are outlined in **Section 3.2**.

### 2.2.3 Multitask Embeddings

Algorithms that integrate different embedding approaches are called multitask embedding models. By combining representations of different data forms, these models greatly enhance embedding's performance on prediction and classification tasks. For example, Yang et al. (2015) use `DeepWalk` and matrix factorization respectively to embed homogeneous graphs with auxiliary text data. Their `text-associated DeepWalk` model achieved state-of-the-art node classification performance and worked very well in sparse and noisy networks with small training samples. Zhang et al. (2017) build on this approach by adding homophily, topology structure, and text data to their embedding model's learning objective function. These data together generate embeddings that better represent the complex interrelations between nodes' local and global structure and the different language contexts over which they exist. While this model also performs well on multi-node classification, it is limited to homogeneous networks and uses a bag-of-words approach to embedding language that does not embed the syntactic and structural information in documents. An et al. (2018) use a similar method to embed networks with auxiliary data that uses triples to distinguish between the different contexts in which two nodes may be related, therefore allowing it to embed heterogeneous knowledge graphs. This model greatly improves the embeddings' quality; however, it too stops at embedding words in nodes' textual context instead of learning to represent the entire text. Xiao et al. (2017) develop a partial solution to this problem of embedding texts in multitask embeddings by averaging texts' word embedding vectors. While this step improves the representations of more complex semantics in texts, it does not embed the larger structural features of documents that may be important to differentiating the language context of nodes.

---

[1] Moreover, it cannot deduce the fact that people from Buffalo, NY, only call them "wings." Homonyms like "Buffalo buffalo Buffalo buffalo buffalo buffalo Buffalo buffalo" are even more complex and out of scope: https://en.wikipedia.org/wiki/Buffalo_buffalo_Buffalo_buffalo_buffalo_buffalo_Buffalo_buffalo.

## 2.3 Suicidality

People who struggle with mental health are often labeled as abnormal and are stigmatized. These labels and stresses of stigmatization, including stereotypes and discrimination, can exacerbate the negative effects of one's mental health status (Link et al. 1989). To cope, many who struggle with mental health avoid discussing it or seeking professional help or social support (Corrigan 2004). Individuals may also withdraw from social life and interaction as they self-stigmatize and try to avoid rejection (Corrigan and Watson 2002). Efforts to reduce mental health stigma find negative attitudes and prejudice are pervasive regardless of age, education, and whether people know others who struggle with mental health (Crisp et al. 2000). Even after such interventions, although levels of stigma decrease among some conditions like depression and anxiety, stigma is still high among the population – especially among teens and young adults (Crisp et al. 2005).

Like mental health illness, suicidality – serious thoughts, plans, and attempts related to suicide – is particularly stigmatized, and epidemiological research finds that 90% of people who attempted suicide had a diagnosable mental health disorder prior to their attempt (Schreiber and Culpepper 2019). This is concerning, as 47,000 individuals in the U.S. commit suicide yearly, making it the $7_{th}$ leading cause of years of life lost (Schreiber and Culpepper 2019). Moreover, these risks are 5-10 times higher for youth and adolescents (Kennebeck and Bonin 2019).

Computational social scientists have studied these dynamics in a variety of contexts. For example, natural language processing has been used to classify depression, post-traumatic stress disorder, attention-deficit/hyperactivity disorder, and other mental health qualities in both clinical and non-clinical settings (Cook et al. 2016; Gkotsis et al. 2017; Gundlapalli et al. 2013; Hollingshead Seitz 2016). Similarly, De Choudhury and Kiciman (2017) find mental health support groups are highly active online and that esteem-boosting comments and network support from other group members lowered one's risk of posting increasingly suicidal submissions in the future. Unfortunately, other work in this setting also finds that significant inequalities persist in regard to which types of posts receive the greatest attention, with dissociative posts receiving the fewest comments and psychotic posts getting the most (Ruch, Kim, and Ruch 2019). Posts presenting with depression and anxiety were neither more nor less likely to receive supportive comments from the community compared to posts labeled with any other mental health condition.

# 3  Data and Methods
## 3.1  Data and Preprocessing

To examine whether graph and language embeddings can be used to predict potentially suicidal individuals in online settings, I used submission data from Reddit – one of the largest social news aggregation, web content rating, and discussion websites. I extracted all public data available on Reddit from June 2005 to June 2018 from pushshift.io, a Reddit-based data project that stores and analyzes Reddit data in real time and releases monthly data dumps of all the information gathered. From this data, I constructed a database of 490 million submissions with 4.3 trillion comments from 66 million authors in 27 million subreddits. Data has a heterogeneous (multimodal) network structure, with authors having links to submissions and comments, comments having links to submissions, and submissions having links to subreddits. All data includes timestamps of when the information was first posted to Reddit, and all data in the database was collected at the moment it was first posted on Reddit – and so all data exists in its original unedited form. The structure of this heterogeneous network is presented in **Figure 1**.

I used a series of Forest Fire network sampling models to collect one main *SuicideWatch* (*SW*) sample and three complementary samples with author seeds from mental health (MH), self-help (SH), and random (R) subreddits[2]. Forest Fire models generate network samples by either starting from a set of seed nodes or random nodes and then "burning" (sampling) every neighboring node of the node set last sampled with a predetermined "burn" probability. If this probability is set to 1, then the Forest Fire model will sample the same as breadth first search. If the "fire" dies out before the predetermined number of nodes is sampled, then the "fire" can restart from an already sampled node and attempt to "reburn" previously unsampled nodes. For sampling large graphs, Leskovec and Faloutsos (2006) demonstrate that the Forest Fire sampling method is one of the most accurate techniques for recovering the "back-in-time" structure of temporal and evolving networks. Kurant et al. (2010), however, note that this method – and every other method based on random walks – is biased toward sampling nodes with large degrees and thus generates samples that overestimate networks' degree. The magnitude of overestimation bias decreases as the fraction of nodes in the network or network subgraph increases. Frank and colleagues (2012), however, find that this bias is relatively low once at least 15% of the target network or subgraph is sampled.

**Figure 1: Heterogeneous Network Structure of Reddit (Comment Nodes Not Shown\*)**

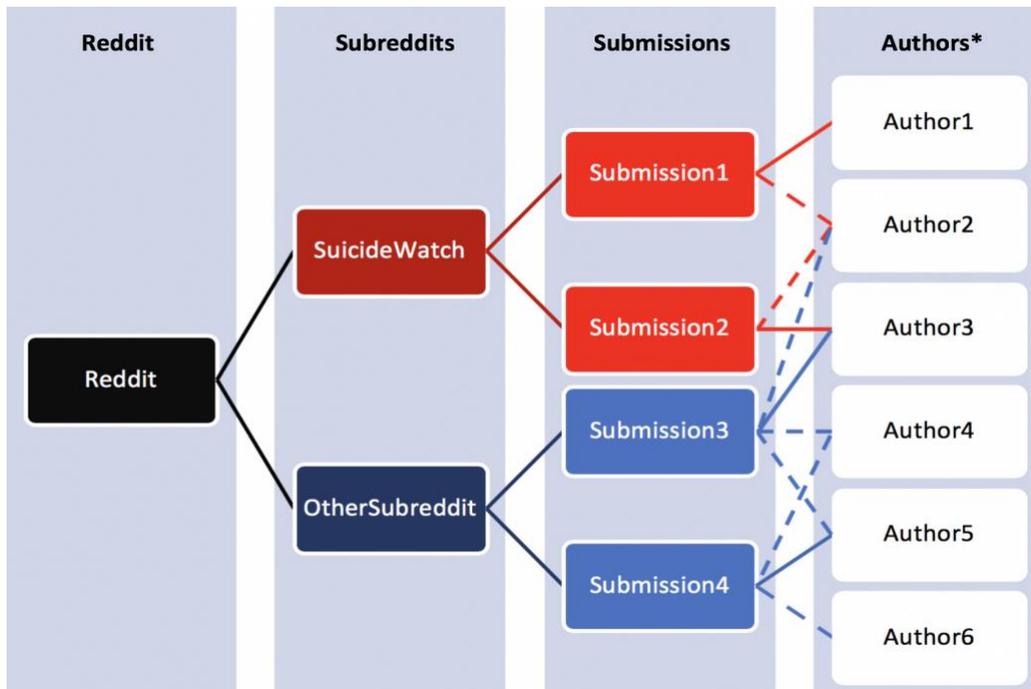

\* For authors, solid lines indicate *direct* submission-author edges (from submission nodes to authors) and dashed lines indicate *indirect* comment-author relations. Comment nodes are hidden; however, they do exist along with comment-author edges (from comment nodes to authors) and submission-comment edges (from submission nodes to comment nodes).

[2] Mental health subreddits include *mentalhealth*, *mentalillness*, *addiction*, *alcoholism*, *Anger*, *Anxiety*, *BipolarReddit*, *depression*, *domesticviolence*, *EatingDisorders*, *OCD*, *Phobia*, *psychoticreddit*, *ptsd*, *schizophrenia*, *survivorsofabuse*, *rape*, *OpiatesRecovery*, *ADHD*, and *itgetsbetter*. Self-help subreddits include *selfimprovement*, *zenhabits*, *productivity*, *personalfinance*, *frugal*, *decidingtobebetter*, *GetMotivated*, *getdisciplined*, *LifeProTips*, and *LifeImprovement*.

The Forest Fire model I constructed sampled nodes as follows. For the main *SW* sample, I first identified all unique authors who posted in *SW* at least once and who had at least 20 submissions in total – regardless of the subreddit in which they posted it. Next, from these 24,281 unique *SW* submission author seeds, I "burnt" 20% of nodes (*V* = 4,948; *V* stands for nodes). I then identified all of these authors' submissions across any subreddit (*V* = 777,243; 94 submissions/author) and "burnt" 20% of them (*V* = 155,646; 31 submissions/author). I identified all of the comments these authors' made (*V* = 9,611,359; 1,942 comments/author) and "burnt" 20% of them (*V* = 1,415,357; 286 comments/author) – making sure to burn no more than 1 comment per unique submission to avoid sampling multiple comments authors may post on the same submission. In this process, I also "burnt" the submission and submission author for which the "burnt" comment was posted. I next identified all comments that were posted to the "burnt" submissions (*V* = 447,579,856) and "burnt" 1% of non-*SW* authors comments (*V* = 4,475,200) and "burnt" all *SW* authors' comments (*V* = 2,109,393). This generated a *SW* sample of 6,584,593 observations (= 4,475,200 + 2,109,393). These observations included data for the following: ~21K subreddits on which ~700K submission authors (=1% of all unique Reddit authors) posted ~1.3M submissions on which ~1.6M comment authors posted ~6.5M comments. Giving the sampling process of the Forest Fire model, the graph produced from these observations generates a connected component.

I then used identically structured Forest Fire models to sample 7.0M observations from a seed list of mental health subreddits, to sample 7.6M observations from a seed list of self-help subreddits, and to sample 5.8M observations from 5,000 randomly selected authors across all subreddits (= the random sample). In this last case, 5000 unique authors were randomly selected as seeds since this number was quite close to the number of unique authors "burnt" by the first "fire" in each of the previous Forest Fire models (e.g., 4,948 unique authors were "burnt" in the *SW* sample). These four contexts were chosen to generate representative samples of severe as well as general mental health related activity in addition to activity related to self-improvement and activity "normal" to Reddit more broadly. Altogether, the four samples generated one large heterogeneous network of 35.5M nodes (*V*) and 190M edges (*E*), which was also a connected component as few of the "burnt" authors had not made at least one submission or comment to one of Reddit's popular subreddits like *r/announcements*, *r/funny*, *r/AskReddit*, and *r/gaming* – the four most followed subreddits.

## 3.2 Graph Embedding with `metapath2vec`

To embed this network while preserving the structural equivalence between nodes, I constructed a metapath-biased random walk sampling algorithm. The walker recursively sampled nodes over metapath context of {subreddit, submission, author, submission, subreddit} with the objective of drawing samples `metapath2vec` (MP2V) could analyze to extract similarities between subreddits through the authors who post submissions in them. Each submission node was walked over 1,000 times for a distance of 100 steps, as recommended by Dong et al. (2017).

Sampling graphs with random walks is computationally efficient in terms of time and memory. First, storing nodes' immediate neighbors is $\mathcal{O}(|E|)$. Retrieving nodes' neighbors is then $\mathcal{O}(|V|)$. Storing interconnections between nodes' neighbors is $\mathcal{O}(a^2|V|)$, where $a$ is the graph's average degree and is usually small for real-world networks (e.g., in the main sample for this study, the graph's average degree is 9.1). Preprocessing transition probabilities makes walking from nodes $\mathcal{O}(1)$. Writing walks' real-time sampling results to disc instead of RAM saves a huge amount of memory. For example, the random walk sample file generated for MP2V was 50 GB; however, a HSBM could not be run on the same network data as it quickly overflowed the server's 128 GB of RAM. Since walks are independent, one can also parallelize the sampler using multiprocessing to greatly enhance sampling speed. This leads MP2V– including sampling and embedding models – to be ~7.6 times more efficient than hierarchical stochastic block modeling while using much less memory (see **Supplemental Note 1**).

The heterogeneous Skip-Gram model MP2V used to learn dense low-dimensional representations of heterogeneous graphs functions as follows. First, heterogeneous graphs are composed of nodes of different types and are described as $G = (V, E, T)$, where $V$ is nodes, $E$ is edges, and $T$ is node type (e.g., author, comment, submission, subreddit). To embed $G$, the heterogeneous Skip-Gram model maximizes the probability of node $v$ having heterogeneous content $N_t(v), t \in T_V$:

$$\arg\max \theta \sum_{v \in V} \sum_{t \in T_V} \sum_{c_t \in N_t(v)} \log p(c_t|v;\theta)$$

where $N_t(v)$ is $v$'s neighborhood among the $t^{th}$ type of nodes, $c_t$ is $v$'s neighboring node of type $t$, and $p(c_t|v;\theta)$ is a softmax function – otherwise known as a multinomial logistic function. The softmax function can be represented as $p(c_t|v;\theta) = \frac{e^{X_{c_t} \cdot X_v}}{\sum_{u \in V} e^{X_u \cdot X_v}}$, where $X_v$ is the $v^{th}$ row of $X$, the embedding vector for node $v$. If the embedding depth for MP2V is 128 dimensions, for instance, then $X_v$ will have 128 elements.

**Figure 2: `metapath2vec` and `metapath2vec++` Skip-Gram Models (from Dong et al. 2017)**

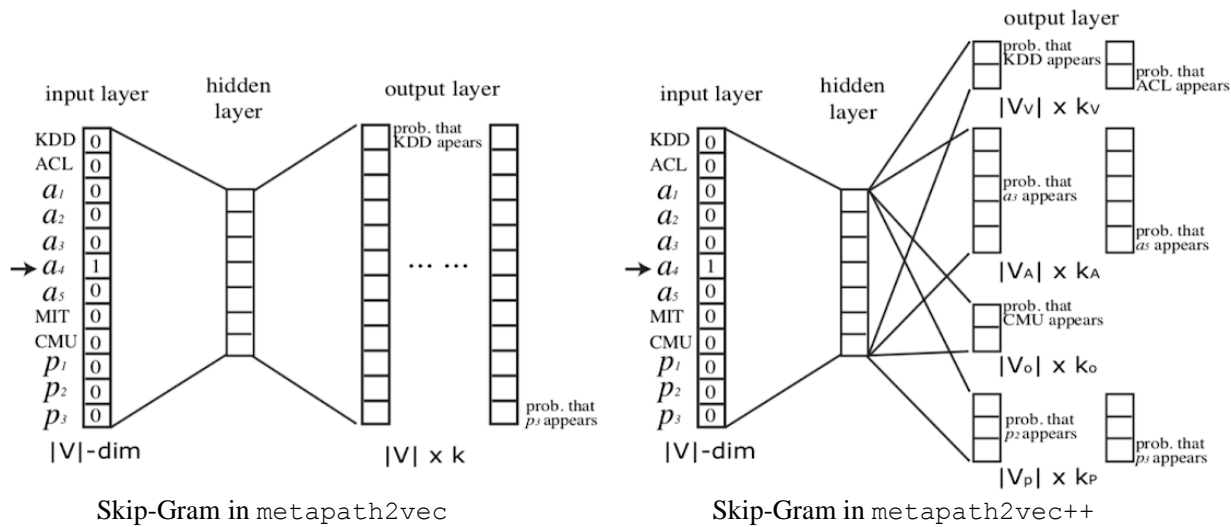

Skip-Gram in `metapath2vec`                    Skip-Gram in `metapath2vec++`

Finally, given the substantial computational complexity involved in calculating the denominator of this softmax function, Dong et al. (2017) modify Mikolov et al.'s (2013) negative sampling procedure to the heterogeneous context, drawing a small sample of size *M* of neighbor and non-neighbor nodes to reduce the number of nodes involved in the denominator. Whereas the MP2V algorithm samples these nodes from the network independent of $v$'s node type, a variant of the `metapath2vec` algorithm called `metapath2vec++` only samples nodes from the same node type as $v$, generating a graph representation in which nodes of the same type are embedded closely to each other and far from nodes of other types; however, because `metapath2vec` has been shown to outperform `metapath2vec++` on many prediction and classification tasks and because the goal of this paper is to predict whether or not an author will post a submission in Reddit's *SuicideWatch* subreddit, I will only use `metapath2vec` here (see **Supplemental Figure 2** for a comparison). In **Figure 2**, the structures of both MP2V variants from (Dong et al. 2017) are presented.

Running MP2V on the full sample of 35.5M nodes and 190M edges over the subreddit, submission, author, submission, subreddit metapath scheme generated embeddings for 1.8M unique subreddit and author nodes. Embeddings had 128 dimensions and were based on a window size of 7 with a negative sampling size of 5. This size is fewer than 35.5M nodes as nodes belonging to submission and comment types were not embedded, because of MP2V's minimum appearance threshold, and since of sampling variation that may have bypassed nodes that otherwise would have surpassed MP2V's minimum appearance thresholds. Total sampling and computation time of these processes was less than one day, and the MP2V model file generated for embeddings was only 0.9 GB. All computing was done on a 32-core 4.1 Ghz 1950X Threadripper™ processor with 128 GB RAM. Code to sample metapaths and run MP2V embedding was obtained from the developer's website: https://ericdongyx.github.io/metapath2vec/m2v.html.

### 3.3 Language Embedding with `doc2vec`

Le and Mikolov (2014) developed what's become known as `doc2vec` (D2V) to learn paragraph vector embeddings that represent texts of arbitrary length (e.g., sentences, paragraphs, documents) in dense low-dimensional space. D2V is trained similarly to MP2V; likewise, two variants of D2V exist: Distributed Bag of Words (DBOW) and a Distributed Memory Model (DMM). The structure of these two models is outlined in **Figure 3**. The DBOW variant functions analogously to MP2V, with the model's goal being to maximize the probability of document *D* being constructed of the different words in the corpus's vocabulary. DMM, by contrast, uses documents' paragraph vector in addition to the ordered words in documents to sequentially predict words that appear next in the document. The authors note that this leads DMM to embed the topic of a document by representing what information is missing from it. Whereas the authors find that DMM generally perform better on prediction and classification tasks, others find that DBOW produces more accurate results (Lau and Baldwin 2016). For this reason, I will follow Le and Mikolov's (2014) advice and concatenate the embeddings separately generated by DBOW and DMM for my downstream prediction tasks.

**Figure 3: `doc2vec` DBOW and DMM Skip-Gram Models (from Le and Mikolov 2014)**

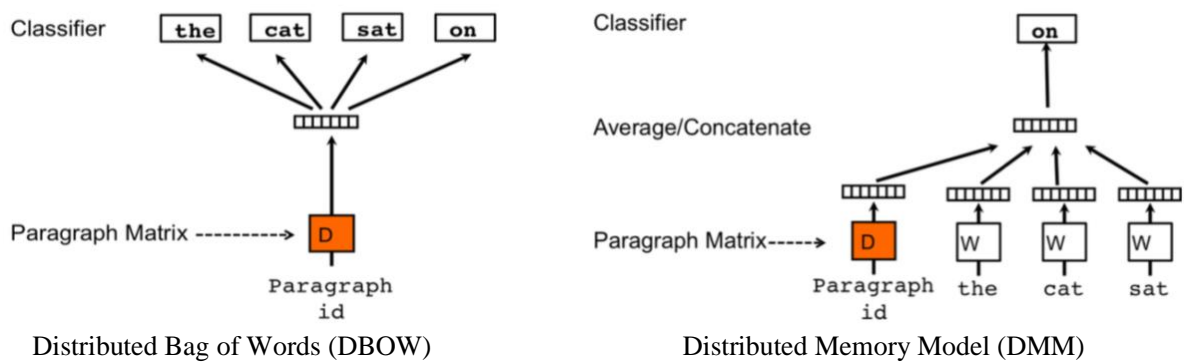

Distributed Bag of Words (DBOW)     Distributed Memory Model (DMM)

I extracted the submission text for all submissions in the main *SW* sample to train my D2V models. This resulted in 1,200,579 documents for training after submissions with no text were excluded (e.g., submissions that posted a photo with no commentary). To preprocess documents for training, I used the `Stanford CoreNLP` toolkit (Manning et al. 2014) and basic Python string methods to lowercase, convert non-ascii characters to similar ascii characters, recode digits as <num> and URLs as <www>, split slash-connected words like "black/white" into "black" and "white", and finally tokenized all text by words. I then used `genism`'s (Řehůřek 2010) implementation of D2V to convert submissions' text into a TaggedDocument object, where each submission is represented as a list of all tokenized words plus document tags for author and subreddit names. This step allows D2V to learn document embeddings for both authors and subreddits. Next, I trained both variants of D2V using a window size of 10, a negative sampling size of 5, and minimum appearance value of 2 – embedding all authors and submissions into 50 dimensions.

I estimated the average and standard deviation of similarities between authors' submissions and *SW* by calculating the inferred DBOW and DMM vectors for each submission authors posted and for *SW,* then calculated each submission's cosine similarity to *SW*, and finally took the average and standard deviation of submissions' similarities to *SW* by author to generate more informative measures of author-level similarities to *SW* – as taking the cosine similarity from authors' D2V embedding and *SW*'s D2V embedding would only provide a measure of similarity without variability. Overall, this generated 689,500 author-level similarities to *SW*. This value does not equal the ~700K submission authors in the main *SW* sample since some authors' submissions were either empty or deleted and therefore such authors were removed before embedding.

## 4 Results

### 4.1 Descriptive Statistics and Embedding Visualizations

**Figure 4: Total Degree Distributions for Samples (via Author-Author Edges)**

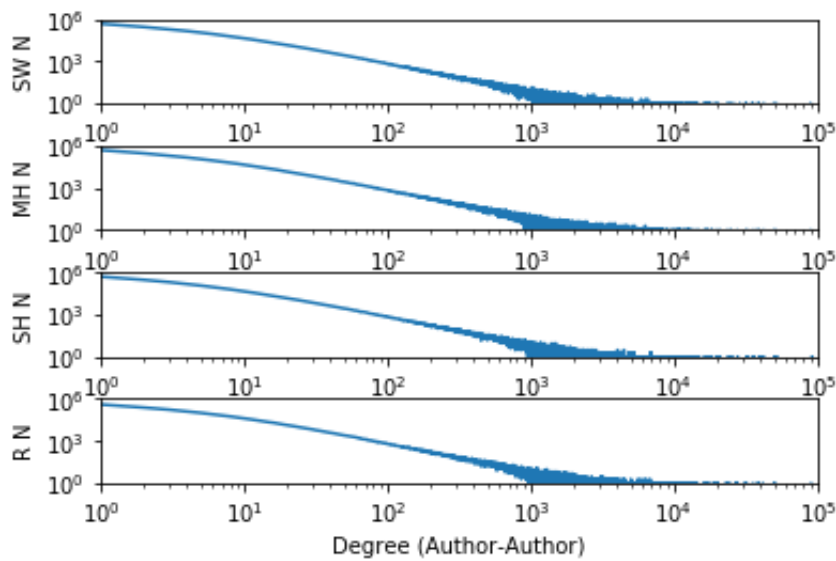

Notes: SW = *SuicideWatch*, MH = mental health, SH = self-help, R = random; author-to-author edges are measured by projecting the hybrid heterogeneous networks to one-mode networks. Total-degree $\bar{x}$ ($\sigma$): SW = 9.1 (0.63), MH = 9.0 (0.59), SH = 9.0 (0.56), R = 9.4 (0.64).

**Figure 5: `UMAP` Visualization of Subreddit-Author-Subreddit `metapath2vec` Embeddings**

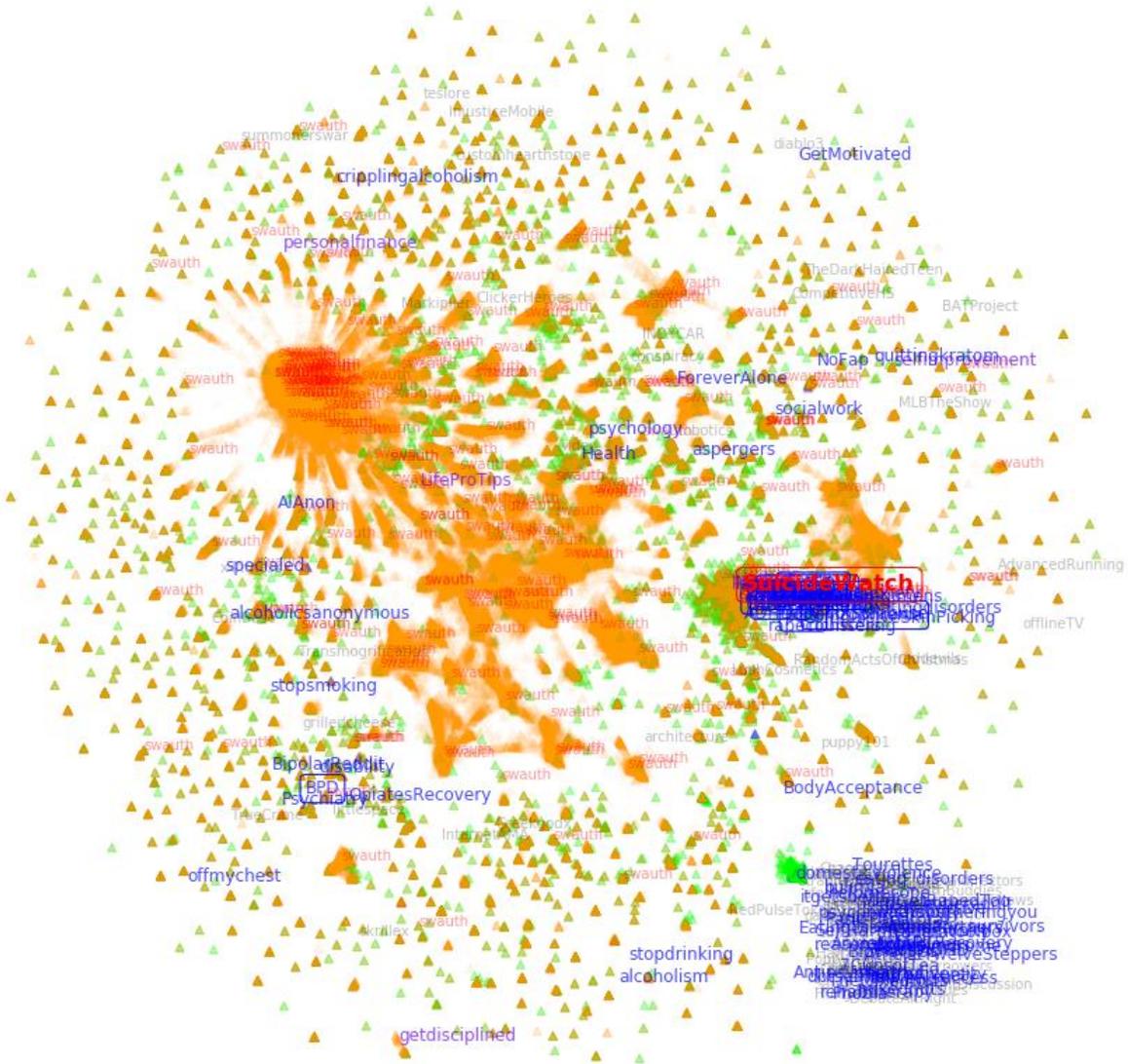

Note: Node annotations include the following: author = orange, *SuicideWatch* = red, mental health subreddit = blue, self-help subreddit = purple, random subreddit = green, transparency = overlapping nodes. Text annotations include the following: *SuicideWatch* subreddit = red with border, *SuicideWatch* author = red, major mental health subreddit = blue with border, minor mental health subreddit = blue, self-help subreddit = purple, random subreddit = gray. All mental health and self-help subreddits are text annotated; however, only 5% of *SuicideWatch* authors (swauth labels in red) and 1% of random subreddits are annotated with text for their names.

**Table 1: Cosine Similarities between *SuicideWatch* and Its 10 Nearest Neighbors**

| | | | | |
|---|---|---|---|---|
| *Depression*: 0.83 | *Anxiety*: 0.82 | *mentalhealth*: 0.75 | *AskDocs*: 0.75 | *MMFB*: 0.74 |
| *Advice*: 0.74 | *socialanxiety*: 0.74 | *selfharm*: 0.73 | *Needafriend*: 0.72 | *StopSelfHarm*: 0.72 |

**Figure 4** shows the total degree distribution for authors' connections (made by collapsing edges from authors to subreddits to other authors into author-to-author edges) among the four subsamples comprising the full sample. Authors have similar levels of connection to other authors across each subsample with very minor differences in their respective averages and standard deviations. This indicates the Forest Fire models that generated the samples drew comparable samples of nodes in which authors from no sample were substantially more or less well-connected to others compared to authors from any other sample. Had such differences existed, they could have added exogenous bias to MP2V's sampling and embedding procedures, as after the four subsamples are combined, highly connected nodes from subsamples with higher total degree distributions would be sampled by the MP2V processes more often than nodes from subsamples with lower distributions. This bias at the very least would affect how often certain nodes and their neighbors are embedded together, which would generate less representative and more imprecise embeddings for such subsamples' nodes. From this point forward, all references to the sample will refer to the full network made of all subsamples combined into one large component.

To visualize the positions author and submission nodes in two-dimensional space, `UMAP` (Uniform Manifold Approximation and Projection (Mcinnes, Healy, and Melville 2018)) was used to reduce the dimension of nodes' 128-dimensional embedding positions. **Figure 5** shows the `UMAP` results. Overall, Figure 5 looks similar to standard social network analysis visualizations (e.g., spring force directed projection) but without edges being drawn between nodes (see **Supplemental Figure 1** to compare MP2V to a spring force directed projection of the author-subreddit projected network and see **Supplemental Figure 2** to compare the MP2V results to those from MP2V++). Various clusters of authors and subreddits are readily visible in **Figure 5**. For example, on the right, *SW* is clustered with *depression, Anxiety,* and self-harm subreddits. A group of bipolar subreddits exists in the lower-left corner. The bottom-right corner has a large cluster of subreddits related to various mental health topics. Finally, the center and top-left corner both have clusters of general subreddits unrelated to mental health or self-help (e.g., *AskReddit*). **Table 1** shows cosine similarities between *SW* and its ten nearest neighbors. All results reflect what one would expect – *SW* is positioned near other mental health subreddits on topics that are highly related to suicidality, including depression, anxiety, self-harm, and social and emotional support (*MMFB* = make me feel better).

**Table 2: Cosine Similarities to *SuicideWatch* among `doc2vec` Models' 15 Nearest Neighbors**

| DMM similarities to *SuicideWatch*: | | DBOW similarities to *SuicideWatch*: | |
| --- | --- | --- | --- |
| **_depression_**, | 0.99 | **_depression_**, | 0.96 |
| *depressed*, | 0.98 | **_MMFB_**, | 0.93 |
| *depression_help*, | 0.98 | *depression_help*, | 0.92 |
| *Suicide_help*, | 0.97 | *whatsbotheringyou*, | 0.92 |
| **_getting_over_it_**, | 0.96 | *depressed*, | 0.92 |
| *Prevent_Suicide*, | 0.96 | *Suicide_help*, | 0.91 |
| **_mentalhealth_**, | 0.96 | **_sad_**, | 0.91 |
| **_sad_**, | 0.96 | *suicidenotes*, | 0.91 |
| **_MMFB_**, | 0.95 | **_offmychest_**, | 0.91 |
| *SanctionedSuicide*, | 0.95 | *SanctionedSuicide*, | 0.90 |
| *venting*, | 0.95 | **_getting_over_it_**, | 0.90 |
| *suicidenotes*, | 0.95 | *venting*, | 0.90 |
| **_mentalillness_**, | 0.95 | **_mentalhealth_**, | 0.89 |
| **_ptsd_**, | 0.94 | **_selfhelp_**, | 0.89 |
| **_BPD_**, | 0.94 | *Vent*, | 0.89 |

Note: Underlined subreddits in bold are also substantially close to *SuicideWatch* in MP2V embedding similarities.

Both variants of the D2V model extracted similar but not overlapping language information from submissions' text. **Table 2** shows DBOW and DMM's similarities in more detail. Overall, there is much similarity among the two language embedding models, but as reflected by subreddit ordering arrows, DBOW and DMM do represent different kinds of language similarities. For example, the top 15 nearest neighbors of *SW* in DMM all have cosine similarities above 0.94, but DBOW similarities range from 0.96 to 0.89. Also, whereas *depression* is the nearest neighbor of *SW* among both models, other subreddits like *MMFB* are higher among the nearest neighbors in DBOW compared to DMM. Many of the nearest neighbors in both D2V variants are also among the nearest neighbors to *SW* in the MP2V node embeddings, noted by subreddits underlined in bold. D2V and MP2V comparisons will discussed more in **Section 4.3**, however.

**Table 3** presents examples of non-*SW* and *SW* submissions taken from submission authors in the main *SW* sample, plus DBOW and DMM cosine similarities to *SW* for submissions. The average and standard deviation of these DBOW and DMM similarity scores for the two sets of submissions (beyond what is shown on screen) are presented below their respective tables. A few qualitative observations on how DBOW and DMM models track different kinds of language information can be drawn from these submissions and their similarities. First, the *dating_tips* submission in the non-*SuicideWatch* table has a high DBOW similarity but low DMM similarity, possibly because it has words that are common in *SW* submissions (e.g., rejection from a loved one) but excludes suicidal topics. Similarly, the top-most submission in the *SW* table has a high DBOW similarity but low DMM similarity, possibly because the topic is on the author's friend. Finally, reoccurring thoughts is a strong indicator of suicidality, which may explain why the bottom-most submission in the *SW* table scores high for both DBOW and DMM. **Table 4** shows the average and standard deviation of DBOW and DMM cosine similarity scores between submissions to other subreddits compared to *SW*. Overall, across all subreddits in the *SW* sample, submissions have relatively low cosine similarity to *SW*, yet submissions to *Anxiety*, *depression*, and *SW* have increasingly similar submissions on average to *SW*, which indicates the two D2V models do distinguish between words and topics that are common in *SW* versus other subreddits.

**Table 3: Examples of Non-*SuicideWatch* and *SuicideWatch* Submissions and their `doc2vec` Cosine Similarities to *SuicideWatch*'s Subreddit-level `doc2vec` Embedding Position**

| Non-*SuicideWatch* subreddit submissions | | | | *SuicideWatch* submissions | | | |
| --- | --- | --- | --- | --- | --- | --- | --- |
| **Subreddit** | **Submission** | **DBOW Dist** | **DMM Dist** | **Subreddit** | **Submission** | **DBOW Dist** | **DMM Dist** |
| *video_game* | here is a guide to win the … | 0.26 | 0.17 | *SuicideWatch* | i have a friend who tried to … | 0.72 | 0.42 |
| *need_advice* | what really bothers me is … | 0.53 | 0.47 | *SuicideWatch* | i want to die and I don't care if … | 0.74 | 0.50 |
| *dating_tips* | rejected by the girl I love … | 0.71 | 0.42 | *SuicideWatch* | keep having reoccurring thoughts … | 0.78 | 0.59 |

$\bar{x}$ ($\sigma$): DBOW = 0.45 (0.15); DMM = 0.40 (0.09)   $\bar{x}$ ($\sigma$): DBOW = 0.77 (0.07); DMM = 0.49 (0.06)

Note: Submission text has been altered so authors cannot be identified. Subreddit names in non-*SuicideWatch* subreddits are also adjusted to be similar to the theme of the original subreddit source.

**Table 4: Average ($\sigma$) `doc2vec` Cosine Similarity Scores to *SuicideWatch* among Submissions to Different Sets of Subreddits:**

| Overall: | *Anxiety:* | *depression:* | *SuicideWatch:* |
| --- | --- | --- | --- |
| DBOW 0.45 (0.15) | DBOW 0.66 (0.09) | DBOW 0.73 (0.08) | DBOW 0.77 (0.07) |
| DMM  0.40 (0.09) | DMM  0.47 (0.05) | DMM  0.48 (0.05) | DMM  0.49 (0.06) |

## 4.2 Embedding Model Comparisons

In **Table 5**, I show the Pearson correlation matrix of embedding similarities to *SW*. Overall, DBOW and DMM embeddings have a moderately strong correlation with each other; however, they are not perfectly overlapping – quantitatively indicating how the two D2V variants learn to represent different kinds of language information in submissions. The MP2V similarities are only slightly correlated with the two D2V similarities, which bodes well for avoiding multicollinearity in the prediction models while also giving evidence that the two forms of embedding models are tracking different kinds information about individuals' behavior and communication and are not extracting redundant similarity data. **Figure 6** shows scatterplots and histograms for the distributions between these similarities to visualize the shape of these correlations. All results are either circular or oval, indicating that heteroskedasticity will not bias the prediction models.

**Table 5: Pearson Correlation Matrix of Embedding Similarities to *SuicideWatch***

|      | MP2V | DBOW | DMM  | D2V$\bar{x}$ |
|------|------|------|------|------|
| **MP2V** | 1.00 | 0.23 | 0.15 | 0.22 |
| **DBOW** | 0.23 | 1.00 | 0.59 | 0.93 |
| **DMM**  | 0.15 | 0.59 | 1.00 | 0.83 |
| **D2V$\bar{x}$** | 0.22 | 0.93 | 0.83 | 1.00 |

Note: D2V$\bar{x}$ = the average of DBOW and DMM embedding similarities.

**Figure 6: Scatterplots and Histograms of `metapath2vec` and `doc2vec` Embedding Similarities to *SuicideWatch***

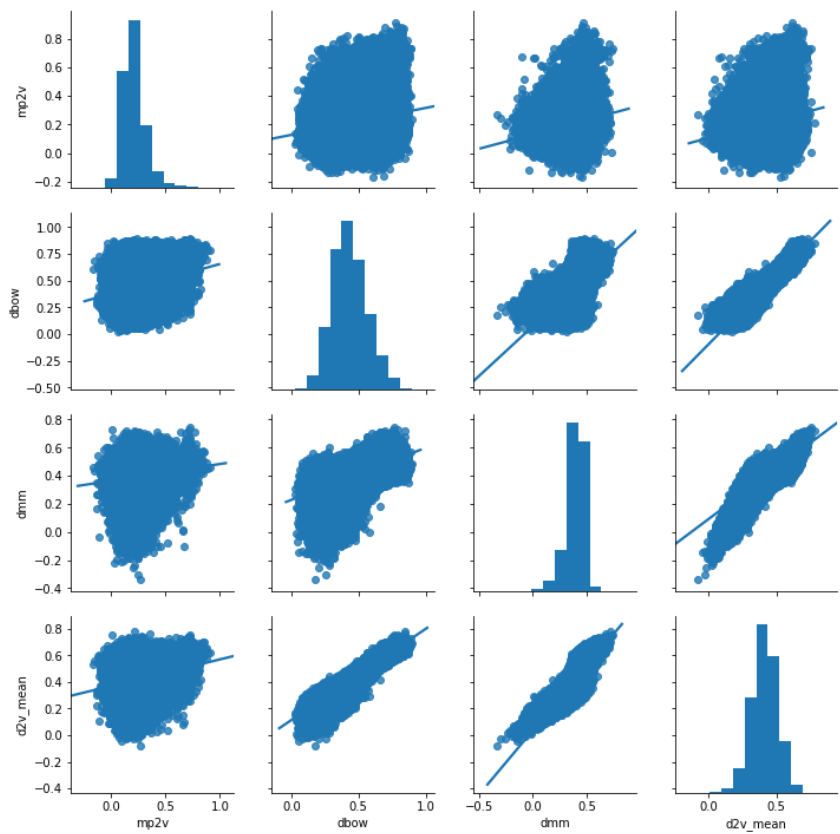

## 4.3 Prediction Tasks

For the prediction tasks, I first truncated training/testing data to 305K by filtering on submission authors from the main *SW* sample and then filtered the training/testing data to 293.5K after any authors who did not have MP2V embeddings were dropped. To balance the training set, I randomly subsampled 4,550 non-*SW* authors as negative non-*SW* cases to compare with 4,060 *SW* authors. For the test set, I drew a random subset of 1,015 *SW* authors and 1,138 non-*SW* authors. I then used a logistic regression model with only the 128-dimension MP2V embeddings as features to predict whether an author will post in *SW*. Testing accuracy (69%) was well above random guess (50% per cell, given the balanced training); however, there were quite a few false-positives and false-negatives (25% and 38% respectively). Overall, however, this model performed quite well for having only included completely unsupervised positional data based on network behavior.

**Figure 7: Confusion Matrices of MP2V-only Prediction Results (Accuracy = 69%)**

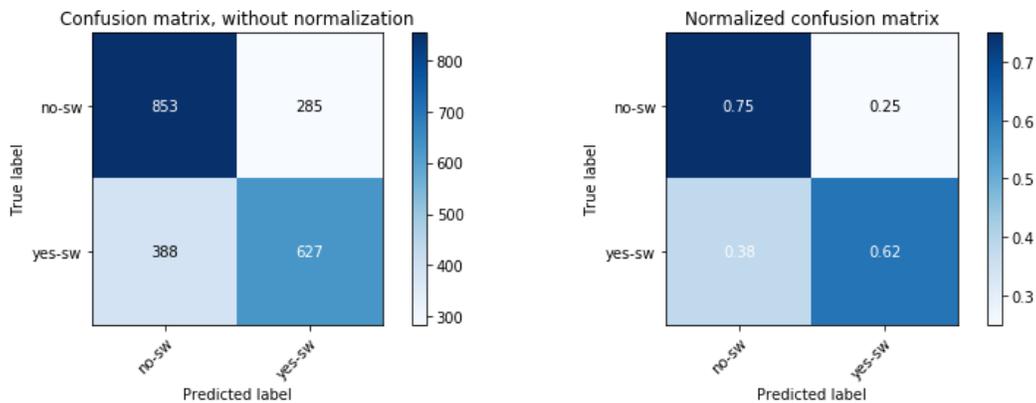

Next, I ran the same prediction model for whether or not a user would post in *SW* using the same training/testing split as above but with only the two D2V models' *SW* distances (i.e., with only 2 covariates instead of 128). Testing accuracy was still quite good at 76%. The predictive language embedding model generated slightly fewer false-positives (21%) than the MP2V-only prediction model but still produced many false-negatives (27%). Overall, however, the model did remarkably well given that it only used two simple embedding distance measures.

**Figure 8: Confusion Matrices of D2V-only Prediction Results (Accuracy = 76%)**

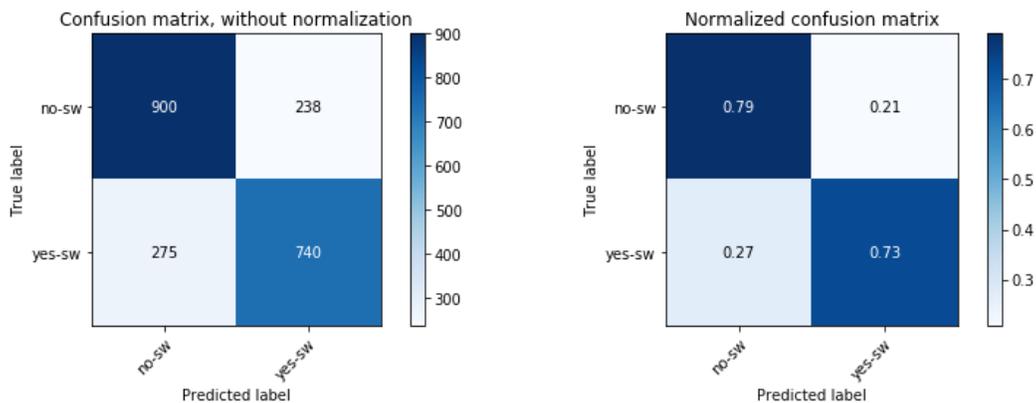

Finally, to test how well one can predict if a user will post in *SW*, I integrated MP2V embeddings and both sets of D2V embedding distances. In addition, I tested this prediction model against all remaining 283K observations without subsampling for the test set to assess how well the model could perform across the full sample (and thus how it may generalize to its analogous population). Subsampling was still used for training to balance the model's learning, however, and so the training set had 4,073 *SW* authors and 5,792 non-*SW* authors while the test set had 1,002 *SW* authors and 282K non-*SW* authors. When used together, integrated MP2V and D2V embeddings predict with high accuracy (90%), precision (89%), and recall (88%) – generating relatively few false-positives and false-negatives (10% and 12%, respectively).

The improved performance of the integrated MP2V+D2V model makes intuitive sense, given the previous results I discussed relating to similarity differences between D2V's DBOW and DMM models and how both variants produced different sets of nearest neighbors from each other and from MP2V. Together, they may each be learning to represent the topics people are talking about, how people are talking about those topics, and where in the network they talk about different topics. In other words, authors' behavior and their language are both important for predicting if authors will post in a particular subreddit – especially for reducing false-positives and false-negatives when target outcomes are rare in a dataset.

**Figure 9: Confusion Matrices of Integrated MP2V-D2V Prediction Results (Accuracy = 90%)**

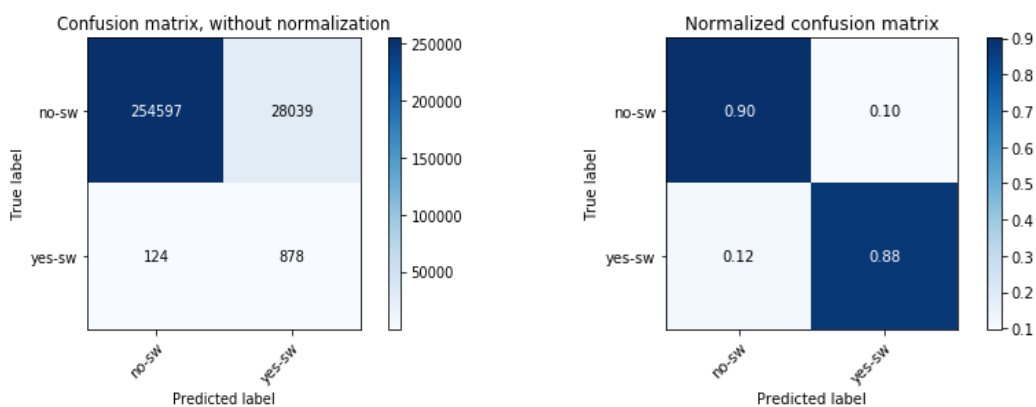

## 4.4 Sensitivity Analysis

To qualitatively evaluate how well the integrated prediction model was able to predict if authors would post in *SW*, I used `UMAP` to generate a two-dimensional projection of the MP2V embeddings of authors and subreddits and annotated accurately predicted *SW* and non-*SW* authors. **Figure 10** shows these results. Even when their MP2V embedding positions were located quite some distance away from *SW* and when they were embedded close to other general subreddits like *AskReddit* in the upper-left corner of the graph, the integrated model was able to accurately predict that these authors would post in *SW*. Similarly, even if authors are located very close to *SW*, the integrated model was able to predict they would be non-*SW* authors. The red and green heatmaps and density plots on the right side of **Figure 10** show these results in greater detail.

**Figure 10: `UMAP` Visualization of Subreddit-Author-Subreddit `metapath2vec` Embeddings and Kernel Density Heatmaps for Integrated MP2V-D2V Model's Predictions of *SuicideWatch* Authors.**

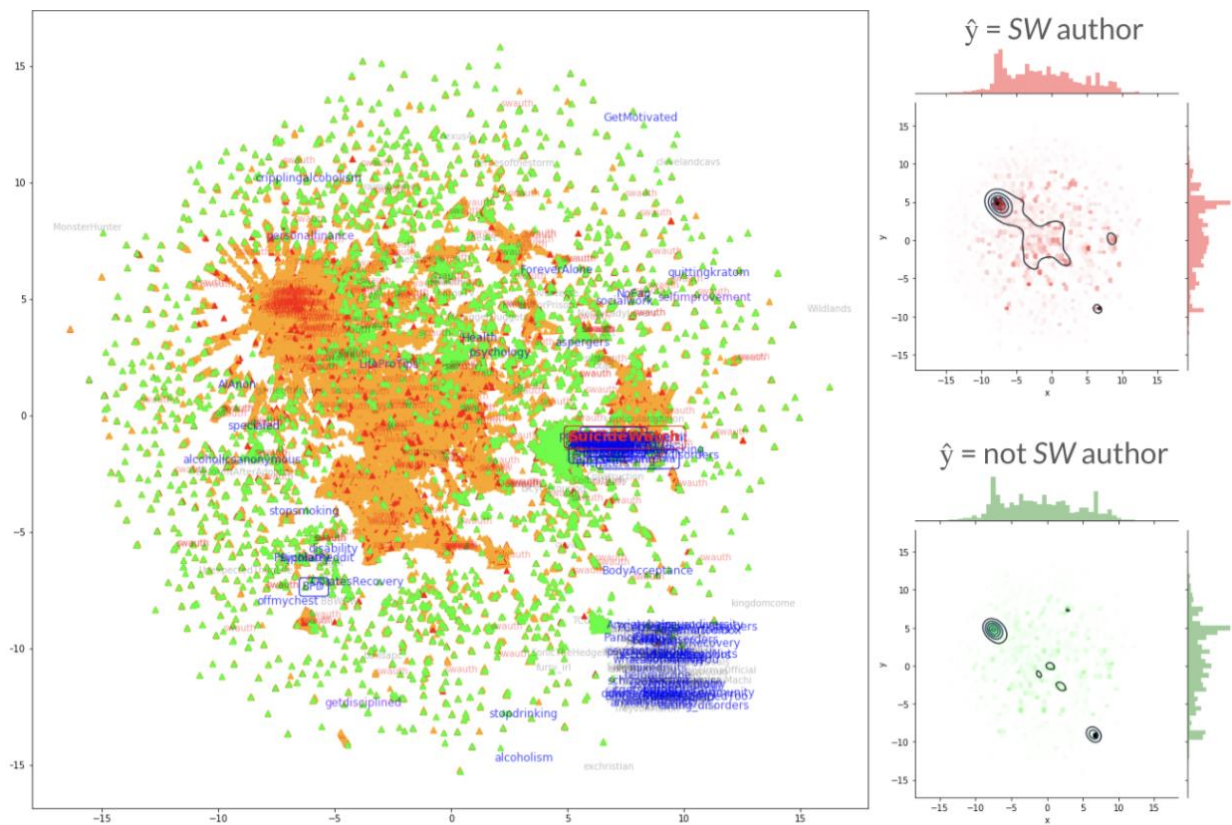

Note: True-positive predicted *SuicideWatch* authors are highlighted in red, and true-negative predicted non-*SuicideWatch* authors are highlighted in green.

## 5 Conclusion

This paper evaluated whether integrating graph and language embedding models (`metapath2vec` and `doc2vec`) could avoid the computational complexity limitations of other popular methods for extracting graph and auxiliary data similarities (e.g., hierarchical stochastic block modeling) from large sparse datasets with the goal of predicting rare events. Empirically, the context for this test was a heterogeneous network of 35.5 million nodes with 190 million edges from Reddit with the objective of predicting whether an author will post a submission in its *r/SuicideWatch* subreddit. In the process, I estimated that MP2V is ~7.6 times more computationally efficient than these other models and required much less memory to run. With no domain expertise needed to embed the heterogeneous network's node and text data, MP2V and D2V models learned similar but qualitatively and quantitatively different embeddings of the data, indicating the two forms of embedding models are in fact learning to represent non-redundant information (e.g., common words, topics, and locations for interacting). Finally, using the embeddings to predict whether an author would post a submission in *SuicideWatch* showed how MP2V and D2V produce better than random predictions when used separately; however, the integrated prediction model using MP2V and D2V together generated highly accurate results with relatively few false-positive and false-negative predictions.

The results of this study extend research on the importance of simultaneously analyzing behavior and language in massive networks and efforts to integrate different kinds of embedding models for prediction and classification tasks, especially when they involve rare events. This is particularly important to emphasize given an ensemble of models – including deep neural networks, boosting, bagging, and stacking – only improved the integrated predictive model's accuracy to 93%. The fact that the ensemble improved the integrated model's performance so little can also be seen as evidence for how embedding models produce high quality dense low-dimensional representations of the large and complex sparse dataset they learn to embed. Though other papers have argued that simply concatenating embeddings from different models may lead to suboptimal results (Zhang et al. 2017), the computational complexity saved by this simpler model may make the tradeoff worth the difference. Given the accuracy of results produced in this context of predicting rare events, the tradeoff seems worthwhile.

Moving forward, fruitful avenues of research following from this work include using the integrated model to better test when and where one form of embedding helps improve performance more than using only one form of embedding model, understanding differences in membership among similar subreddits (e.g., comparing *alcoholicsanonymous*, *AlAnon*, *cripplingalcoholism*, *stopdrinking*, and *addiction*), and finally to predict the presentation of psychiatric attributes in text (e.g., suicidality, depression, anxiety, substance abuse, etc.). A methodological contribution that integrated graph and language embedding models could make is in analyzing and predicting social influence, social contagion, and other social dynamics. For example, Qiu et al. (2018) use graph embeddings to test neighbors' social influence on individuals' decisions to take different actions over time. Integrating language embeddings into this framework could help differentiate between individuals who share connections with others but who are different from them in cultural or ideological ways that are not reflected in network ties alone (e.g., a Twitter user who replies to a politician's Tweet but does so to critique the politician's message, stance, or political party).

The application-specific results of this study also have important clinical and social relevance for mental health research, social support efforts, and interventions. First, I emphasized the importance of drawing representative network samples to improve models' generalizability. Many studies of mental health dynamics on social media draw simple random samples of individuals; however, by foregoing a network-based sample, such studies lose the ability to directly incorporate information about how one's network behavior, connectivity, and neighborhood effects may shape their results. Positive and negative social influence from supportive and potentially harmful network neighbors and network environments (e.g., hopeful versus despairing attitudes) are key predictors of whether individuals' mental health will improve or decline, and network-based samples are able to extract this kind of structural and relational data whereas simple random samples extract information on individuals as if they exist in isolation from one another. Studies testing social support can more accurately capture such dynamics through network-based samples.

Secondly, demonstrating that integrating graph and language data generates better predictions of suicidality aligns with clinical psychiatric diagnostic criteria which includes behavioral and verbal indicators of suicidality. These professionals use both forms of information to corroborate each other and avoid false-positive and false-negative diagnoses (e.g., in cases where one says they are not suicidal but has intentionally overdosed on medication). Lastly, predicting individuals at risk of becoming suicidal may allow for early interventions to help those at risk. Facebook is presently developing such systems in partnership with the Crisis Text Line, the National Eating Disorder Association, and the National Suicide Prevention Lifeline. It is important to note, however, that such interventions may be detrimental to one's self-esteem should they be inaccurate or unwanted (e.g., in cases where one is in fact receiving professional help and wishes to have privacy). Mental health is a sensitive topic, making those who struggle with it vulnerable. Any effort to implement system-wide screening and intervention systems, whether algorithmically- or user-driven, should be acceptable and welcomed by the given community. Otherwise, one may risk destroying trust and confidentiality in the group and cause people to leave it – defeating the purpose of social support communities and potentially causing further harm to individuals' well-being.

While the results of this study are promising in many ways, they are qualified by a few limitations. First, neither the graph embeddings nor the language embeddings presented here are dynamically constructed. In other words, neither embedding learns to represent information on how individuals' interaction or communication patterns evolve over time. Such data are important for predictions, as they can indicate whether one is moving towards or away from an outcome of interest. Some of the individuals predicted to be at high risk of suicidality in this study may have been at risk when they first joined Reddit but then moved farther from risk over time, whereas others may have had little risk when joining Reddit but then became increasingly at risk as time passed. Second, being at risk of suicidality in this study is defined as posting a submission in *SuicideWatch*. Not everyone who is suicidal may post in *SuicideWatch* (e.g., selection biases)*,* and an estimated 22% of users who do post in *SuicideWatch* have no significant evidence of suicidality when their submissions are evaluated by a physician specializing in psychiatry (Ruch, Kim, and Ruch 2019). Some users may also post in *SuicideWatch* with throwaway or secondary accounts. For example, some of the users in this study had words related to "throwaway" in their name. Attempts to avoid such users was made by avoiding sampling users who had fewer than 20 submissions over their lifetime, but this method does not avoid problems when someone uses a throwaway or secondary account many times. Overall, these accounts would limit attempts to sample the true users' complete network of interaction and language events and could lead to such users being sampled twice (i.e., a primary and secondary account are both sampled for one user). Lastly, since online contexts have different norms, incentives, and constraints than offline contexts, the integrated model presented here may not generalize to other contexts. Reddit, for example, provides more anonymity than Facebook, which may affect how people present themselves and interact with others.

# 6   Acknowledgements

This work was supported in part by NSF grant SES-1756822 and NIH grant R25HD079352. The views and conclusions presented in this study are those of the author and should not be interpreted as necessarily representing the official policies or endorsements – either expressed or implied – of NSF, NIH, the U.S. Government, Cornell University, or Graphika Inc. The author wishes to thank Drs. Jennifer Ruch, Michael Macy, David Mimno, Thomas Gilovich, Vladimir Barash, James Moody, and Christopher Bail for feedback and support on parts of this project. The author is also grateful for assistance and feedback from Seunghyun Kim, Lillyan Pan, Hannah Lee, James Zou, Jeffrey Tsang, Bryan Min, Juliana Hong, and Cornell University's Social Dynamics Laboratory.

# Supplemental Materials

**Supplemental Note 1: Estimated Time Complexity of Clustering and Embedding Functions**

SBM: $O(V \ln_2 V) \rightarrow O(SMB_{V=10M}) = 2.6e9$

HSBM: $O(V \ln_2 V \times Community\_Shrinking \times Hierarchies) \rightarrow O(HSBM_{V=10M, Hier=10}) = 2.6e16$, where $Community\_Shrinking = O(Average\_Blocks_2)$ and $Average\_Blocks = 1000$

MP2V: $O((V\_seeds \times walks \times walk\_length) + (V\_sampled \times Neg\_samp \times mp2v\_iter)) \rightarrow O(MP2V_{V=10M, Iter=30}) = 3.4e15$, where $V\_sampled = 1.8M$ and $Neg\_samp = O(V\_seeds \times \log(V\_sampled))$

∴ `metapath2vec` is ~7.6 times more time efficient than hierarchical stochastic block models.

Notes: SMB = stochastic block model; HSBM = hierarchical stochastic block model; MP2V = `metapath2vec`. Time complexity estimates are based on [graph-tool.skewed.de/static/doc/inference.html#](graph-tool.skewed.de/static/doc/inference.html#), [https://pdfs.semanticscholar.org/fb95/6aead82fb3715b2910805b0248526cf7e428.pdf](https://pdfs.semanticscholar.org/fb95/6aead82fb3715b2910805b0248526cf7e428.pdf), [https://stackoverflow.com/questions/54950481/word2vec-time-complexity](https://stackoverflow.com/questions/54950481/word2vec-time-complexity) (given that `metapath2vec`'s time complexity is approximately equal to that of `word2vec`).

**Supplemental Figure 1: Comparison of Traditional Network Position Visualization (using Spring Force Directed Projection) to `metapath2vec` Embedding Position Projection (using `UMAP`)**

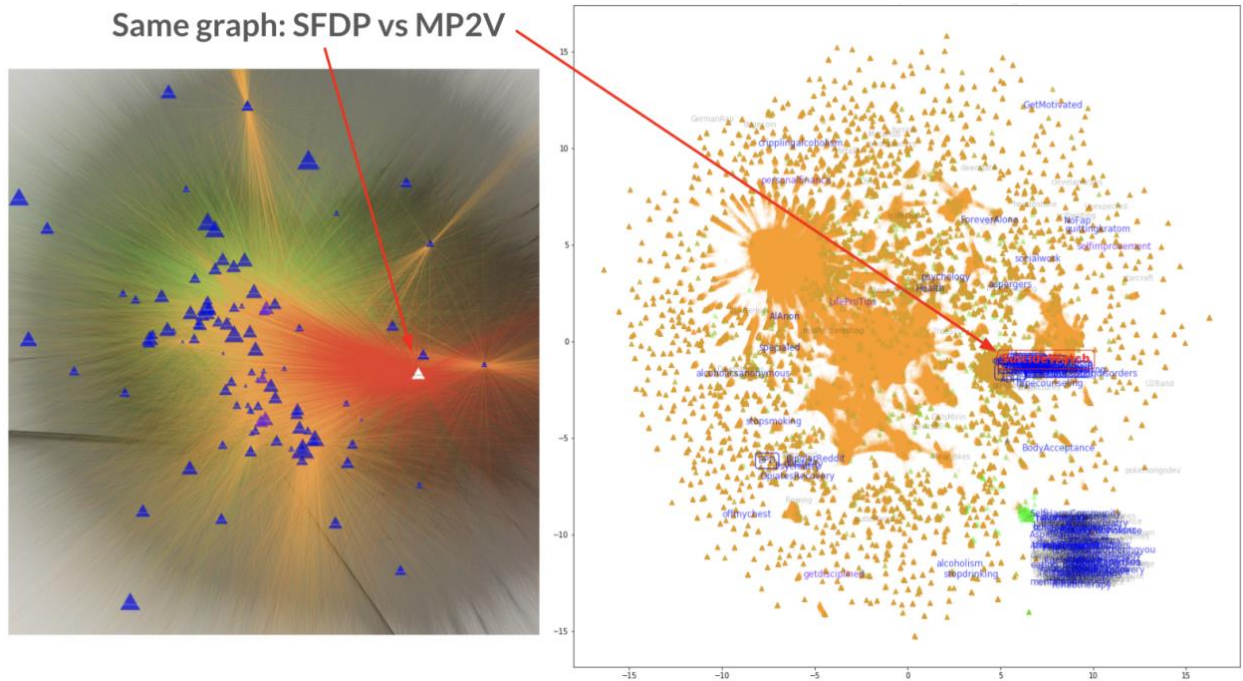

Notes:
 - *SuicideWatch* node = white, edges to *SuicideWatch* node = red
 - Mental health nodes = blue, edges to mental health nodes = orange
 - Self-help nodes = purple, edges to self-help nodes = green
 - Random nodes = green, edges to random nodes = gray
 - Node size = node degree

**Supplemental Figure 2: `UMAP` Projection of Subreddit-Author-Subreddit `metapath2vec++` Embeddings**

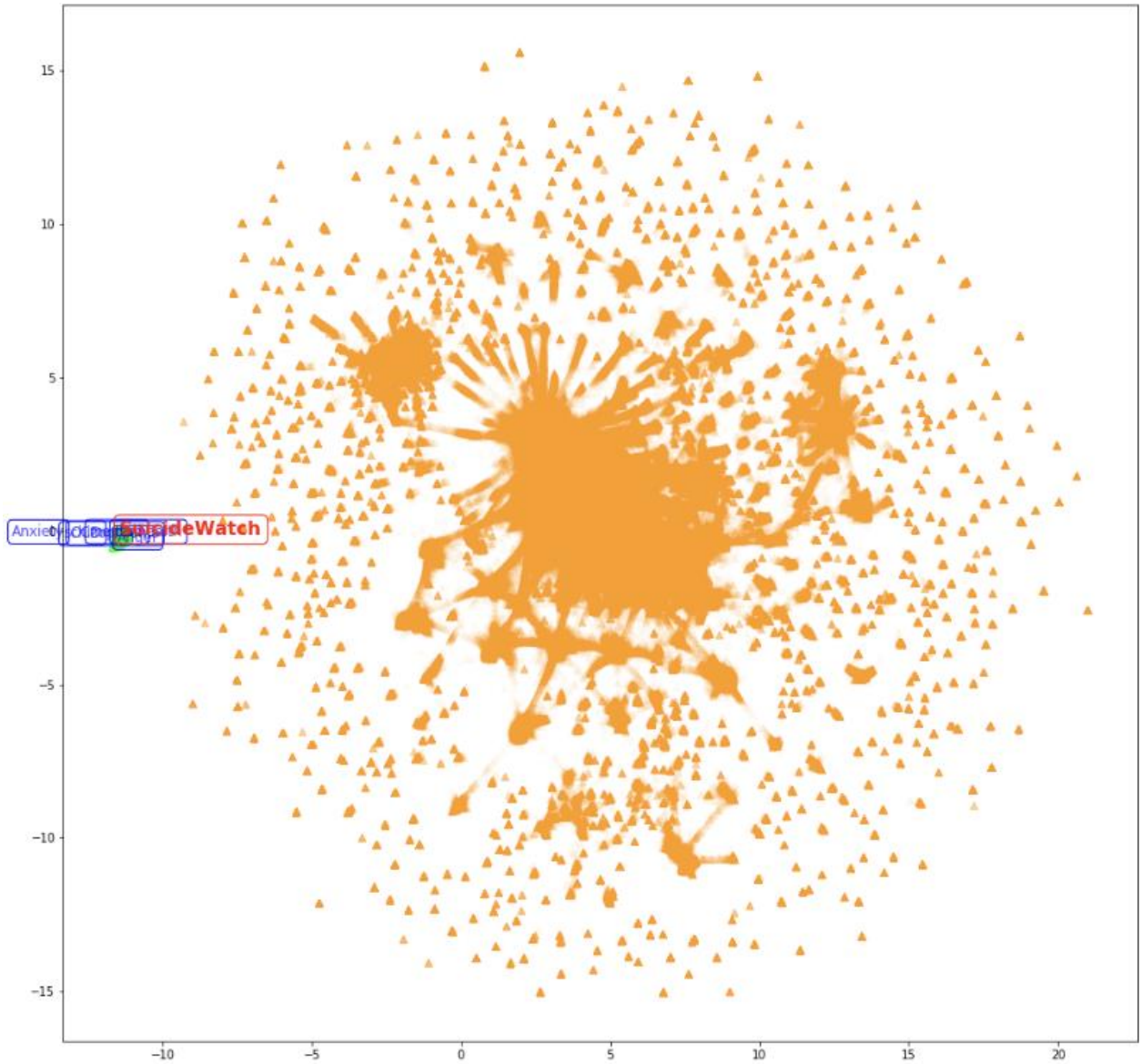

Note: The top nearest neighbors of *SW* in MP2V++ have very high overlap with those of *SW* in MP2V. For example, compare the results of **Table 1** to MP2V++'s nearest neighbors shown below. All top-10 neighbors of *SW* in MP2V are among the top-15 nearest neighbors of *SW* in MP2V++.

**Cosine Similarities between *SW* and Its 10 Nearest Neighbors via `metapath2vec++`**

| | | | | |
|---|---|---|---|---|
| *depression*: 0.90 | *selfharm*: 0.86 | *Anxiety*: 0.85 | *Advice*: 0.85 | *socialskills*: 0.82 |
| *BPD*: 0.82 | *MMFB*: 0.82 | *ForeverAlone*: 0.82 | *offmychest*: 0.81 | *mentalhealth*: 0.80 |